
\magnification 1200
\hsize=6.0 true in
\hoffset=0.0 true in
\baselineskip 15pt
\def\KK{K^0-\bar{K^0}}
\def\BB{B^0-\bar{B^0}}
\def\DM{\Delta M}
\def\lsim{\ ^<\llap{$_\sim$}\ }

\def\r2{\sqrt 2}
\def\ra{r_\alpha}
\def\rb{r_\beta}
\def\rmuu{\gamma^{\mu}}
\def\rmud{\gamma_{\mu}}
\def\PL{{1-\gamma_5\over 2}}
\def\PR{{1+\gamma_5\over 2}}
\def\sw2{\sin^2\theta_W}
\def\v#1{v_#1}
\def\tb{\tan\beta}
\def\sb{\sin\beta}
\def\cb{\cos\beta}
\def\s2b{\sin 2\beta}
\def\c2b{\cos 2\beta}
\def\s2b2{\sin^22\beta}
\def\uL{{\tilde u}_L}
\def\uR{{\tilde u}_R}
\def\cL{{\tilde c}_L}
\def\cR{{\tilde c}_R}
\def\tL{{\tilde t}_L}
\def\tR{{\tilde t}_R}
\def\MCH{M_{H\pm}}
\def\mgr{m_{3/2}}
\def\m#1{{\tilde m}_#1}
\def\mH{m_H}
\def\mwi{{\tilde m}_{\omega i}}
\def\mw#1{{\tilde m}_{\omega #1}}
\def\mQ{{\tilde m}_Q}
\def\mU{{\tilde m}_U}
\def\MuSL{{\tilde M}_{uL}^2}
\def\MuSR{{\tilde M}_{uR}^2}
\def\McSL{{\tilde M}_{cL}^2}
\def\McSR{{\tilde M}_{cR}^2}
\def\MtSk{{\tilde M}_{tk}^2}
\def\MtS#1{{\tilde M}_{t#1}^2}
%
\line{\hfill CERN-TH.7345/94}
\line{\hfill TKU-HEP 94/04}
\line{\hfill CFIF-UTL/IST 2/94}
\  \
\vskip 1.0 true cm
\centerline{\bf
Supersymmetric Contributions to $\BB$ and $\KK$ Mixings}
\vskip 2.0 true cm
\centerline{G. C. Branco$^1$\footnote{$^*$}{On leave from
Departamento de F\'isica and CFIF/UTL, Instituto Superior
T\'ecnico, Av. Rovisco Pais, 1096 Lisboa Codex, Portugal.},
G. C. Cho$^2$\footnote{$^\dagger$}{Research Fellow of the Japan Society
for the Promotion of Science}, Y. Kizukuri$^2$, and N. Oshimo$^3$}
\vskip 0.5 true cm
\centerline{$^1$\it CERN, Theory Division}
\centerline{\it CH-1211 Gen\`eve 23, Switzerland}
\centerline{$^2$\it Department of Physics, Tokai University}
\centerline{\it 1117 Kita-Kaname, Hiratsuka 259-12, Japan}
\centerline{$^3$\it GTAE/CFIF, Instituto Superior T\'ecnico}
\centerline{\it Av. Rovisco Pais, 1096 Lisboa Codex, Portugal}
\vskip 2.0 true cm

\centerline{\bf Abstract}
\vskip 0.5 true cm
     We show that in the supersymmetric
standard model (SSM) box diagrams mediated by
charginos and up-type squarks may give large contributions
to $\BB$ and $\KK$ mixings.  This is due to the fact
that a heavy top quark mass allows large mass differences
among the up-type squarks, which leads to a less efficient
cancellation among the different squark diagrams.
If the ratio of the vacuum expectation values of the Higgs
bosons is of order of unity, box diagrams mediated by
charged Higgs bosons also contribute
appreciably.   For sizable regions of the SSM parameter
space, the SSM contributions could be sufficiently
large to affect significantly the evaluation of
the CKM matrix from the experimental values of
the mixing parameter $x_d$ and the $CP$ violation
parameter $\epsilon$.

\smallskip
\vfill\eject

     The supersymmetric standard model (SSM) predicts
new contributions to
flavor changing neutral current (FCNC) processes,
in addition to those already present in the standard
model (SM).  These new contributions arise from
interactions coupling a quark to a squark of a different
generation and a chargino, a neutralino, or a gluino [1].
Since the SSM contains two doublets of Higgs bosons,
there also exist FCNC processes mediated by charged
Higgs bosons [2].
The effects of these new contributions on $\KK$ and $\BB$
mixings, as well as on other FCNC processes, have been
extensively analyzed. (For a review, see ref. [3])
Initially it was thought that the dominant SSM contribution
to the above mixings would arise from
gluino-mediated box diagrams and that it would be substantial.
However, later analyses [4], taking into account all the
new contributions of FCNC, led to the result that
box diagrams mediated by charged Higgs bosons
gave the largest contribution to $\BB$ mixing.
It was also shown that
this charged Higgs boson contribution was at most a
few tenth of the SM contribution, making it difficult
to observe supersymmetric effects through those mixings.

     The recent experimental evidence for a rather
large $t$-quark mass of 174$\pm 10_{-12}^{+13}$ GeV [5]
provides motivation to reconsider the SSM contributions
to $\KK$ and $\BB$ mixings.
In the SSM, such a large $t$-quark mass theoretically
implies the possible existence of a light $t$-squark,
provided the squark masses of the first two
generations are around 200 GeV or less.
Lack of experimental evidence for supersymmetric particles
at the Tevatron suggests that the masses
of most of the squarks are larger than 150 GeV,
but leaves open the possibility of having a lighter
$t$-squark [6], the
lower bound on its mass being given only by LEP [7].
There is thus the possibility of having large mass differences
among the up-type squarks.  In this case
the chargino contributions
to $\KK$ and $\BB$ mixings
become potentially
important, since the box diagrams
with different up-type
squarks no longer have the severe cancellation
which is present in the case of almost degenerate squark masses.
In fact, in the radiative
decay $b\rightarrow s\gamma$, it was shown [8] that,
among the new contributions of FCNC, chargino-mediated
diagrams gave the dominant contribution.  In particular,
if one of the $t$-squarks
is much lighter than the other squarks, the chargino
contribution to $b\rightarrow s\gamma$
can become as large as the SM contribution.
In view of the above, the question
naturally arises whether
the existence of a light $t$-squark can also imply
large chargino contributions to $\KK$ and $\BB$ mixings.

     In this letter we study $\BB$ and $\KK$ mixings
within the SSM, in light of the recent
evidence of a heavy $t$-quark, concentrating
our attention on the effects of the
chargino and charged Higgs boson contributions.
These short distance contributions affect
the mass difference $\Delta M_B$ of the $B^0$-mesons
and the $CP$ violation parameter $\epsilon$ in the $K$-meson
decays.   We will show that the
SSM contributions to these quantities
are indeed large for sizable
regions of the SSM parameter space, where a
chargino, a $t$-squark, and/or a  charged Higgs boson
are predicted to have masses to be explored
at the next generation of colliders.  Constraints on
the SSM will also be discussed by confronting the
elements of the Cabibbo-Kobayashi-Maskawa (CKM) matrix
inferred from $\BB$ and $\KK$ mixings with those
measured by other phenomena.

     We assume that the masses and mixings of the
squarks are given by
the model based on $N$=1 supergravity and
grand unification (for reviews, see ref. [9]).
At the electroweak scale
the interaction eigenstate squarks are mixed in
generation space
through mass-squared matrices.  These generation
mixings among the up-type squarks are approximately
removed by the same matrices that diagonalize the mass
matrix of the up-type quarks.  As a result the generation
mixings appearing in
the interactions between down-type quarks and
up-type squarks in the mass eigenstate basis can be described
by the CKM matrix of the quarks.

     In each flavor the left-handed and right-handed
squarks are mixed by the Yukawa interaction.
For the first two generations, however,
these mixings can be neglected
due to the smallness of the corresponding quark masses.  The
masses of the left-handed squarks $\uL, \cL$ and the
right-handed squarks $\uR, \cR$ are given by
$$\eqalign{
\MuSL =\McSL &= {\mQ}^2+\c2b({1\over 2}-{2\over 3}\sw2)M_Z^2, \cr
\MuSR =\McSR &= {\mU}^2+{2\over 3}\c2b\sw2 M_Z^2, \cr
             &\tb = {\v2 \over\v1}, }
\eqno(1)
$$
where $\v1$ and $\v2$ stand for the vacuum expectation values
of the Higgs bosons.  The mass parameters $\mQ$ and $\mU$ are
determined by the gravitino and gaugino masses.
Assuming a common mass at around the grand unification scale,
the difference ($\mQ-\mU$) arises only from the electroweak
interactions, and it is small compared to $\mQ$, $\mU$.
For the third generation, the large
$t$-quark mass $m_t$ leads to an appreciable
mixing between $\tL$ and $\tR$.
The mass-squared matrix for the $t$-squarks
is given by
$$
M_t^2 = \left(\matrix{\MuSL +(1-|c|)m_t^2 &
                     (\cot\beta\mH +a_t^*\mgr)m_t \cr
              (\cot\beta\mH^* +a_t\mgr)m_t &
                \MuSR +(1-2|c|)m_t^2 }    \right),
\eqno (2)
$$
where $\mgr$ and $\mH$ denote the gravitino mass and the
higgsino mass parameter, respectively.  The dimensionless
constants $a_t$ and $c$ depend on other SSM
parameters:  $a_t$ is related to the breaking of local
supersymmetry and its absolute value is constrained
to be less than 3 at the grand unification scale;
$c$ is related to the radiative corrections to the
squark masses.  At present, these values cannot be
theoretically fixed, although it can be asserted that
their magnitudes are of order of unity.
The mass eigenstates of the $t$-squarks are obtained by
diagonalizing the matrix $M_t^2$:
$$
  S_tM_t^2S_t^\dagger ={\rm diag}(\MtS1, \MtS2),
\eqno (3)
$$
where $S_t$ is a unitary matrix.

     The charginos are the mass eigenstates of
the $SU(2)$ charged gauginos and the
charged higgsinos.  Their mass
matrix is given by
$$
    M^- = \left(\matrix{\m2 & -{1\over\r2}g\v1 \cr
                -{1\over\r2}g\v2 & \mH}        \right),
\eqno (4)
$$
where $\m2$ is the $SU(2)$ gaugino mass.  The unitary matrices
$C_L$ and $C_R$ diagonalize this matrix as
$$
      C_R^\dagger M^-C_L = {\rm diag}(\mw1, \mw2).
\eqno (5)
$$
The lagrangian for the chargino$-$squark$-$quark interactions
can then be written
explicitly in terms of particle mass eigenstates [8].

The $\BB$ mixing receives contributions from the
box diagrams in which the charginos and the up-type
squarks are exchanged.  For
the effective lagrangian of the $\Delta B=2$ process
we obtain:
$$\eqalignno{
& L^C_{\Delta B=2} = {1\over 8M_W^2}({g^2\over 4\pi})^2
            \sum_{a,b}\sum_{i,j}\sum_{k,l}
                 V_{a1}^*V_{a3}V_{b1}^*V_{b3}   &(6)\cr
      & [F^C_1(a,k;b,l;i,j)\bar d\rmuu\PL b\bar d\rmud\PL b
         + F^C_2(a,k;b,l;i,j)\bar d\PR b\bar d\PR b],      }
$$
$$\eqalign{
F^C_1(a,k;b,l;i,j) &=
           {1\over 4}G^{(a,k)i}G^{(a,k)j*}G^{(b,l)i*}G^{(b,l)j}
                      Y_1(r_{(a,k)},r_{(b,l)},s_i,s_j),   \cr
F^C_2(a,k;b,l;i,j) &=H^{(a,k)i}G^{(a,k)j*}G^{(b,l)i*}H^{(b,l)j}
                      Y_2(r_{(a,k)},r_{(b,l)},s_i,s_j),   }
$$
where $V$ denotes the CKM matrix,  $a, b$ are
generation indices,
and $i, j$ and $k, l$ respectively stand for the
two charginos and the two squarks in each flavor.
The functions $Y_1, Y_2$ and their arguments
coming from loop integrals are given by
$$\eqalignno{
 Y_1&(\ra,\rb,s_i,s_j)                    \cr
   & =  {\ra^2\over (\rb-\ra)(s_i-\ra)(s_j-\ra)}\ln \ra
     +{\rb^2\over (\ra-\rb)(s_i-\rb)(s_j-\rb)}\ln \rb  \cr
   & +{s_i^2\over (\ra-s_i)(\rb-s_i)(s_j-s_i)}\ln s_i
     +{s_j^2\over (\ra-s_j)(\rb-s_j)(s_i-s_j)}\ln s_j, &(7)\cr
 Y_2&(\ra,\rb,s_i,s_j)                    \cr
   & = \sqrt{s_is_j}[{\ra\over (\rb-\ra)(s_i-\ra)(s_j-\ra)}\ln \ra
     +{\rb\over (\ra-\rb)(s_i-\rb)(s_j-\rb)}\ln \rb  \cr
   & +{s_i\over (\ra-s_i)(\rb-s_i)(s_j-s_i)}\ln s_i
     +{s_j\over (\ra-s_j)(\rb-s_j)(s_i-s_j)}\ln s_j], \cr
   & r_{(1,1)} =r_{(2,1)} ={\MuSL\over M_W^2}, \quad
     r_{(1,2)}=r_{(2,2)}={\MuSR\over M_W^2}, \quad
     r_{(3,k)} ={\MtSk\over M_W^2}, \cr
   & s_i ={\mwi^2\over M_W^2}, }
$$
and the coupling constants $G^{(a,k)i}$, $H^{(a,k)i}$ are:
$$\eqalign{
G^{(1,1)i} &=G^{(2,1)i}=\r2 C_{R1i}^*,  \quad
                            G^{(1,2)i}=G^{(2,2)i}=0,    \cr
G^{(3,k)i} &=\r2 C_{R1i}^*S_{tk1}-{C_{R2i}^*S_{tk2}\over\sb}
                                          {m_t\over M_W}, \cr
H^{(1,1)i} &=H^{(2,1)i}={C_{L2i}^*\over\cb}{m_b\over M_W}, \quad
                             H^{(1,2)i}=H^{(2,2)i}=0,    \cr
H^{(3,k)i} &={C_{L2i}^*S_{tk1}\over\cb}{m_b\over M_W}. }
\eqno (8)
$$
In eq. (6) each term is proportional to $V_{11}^*V_{13}$,
$V_{21}^*V_{23}$, or $V_{31}^*V_{33}$.  However, from
unitarity of the CKM matrix, the following relation
holds:
$$\eqalignno{
\sum_{a,b}V_{a1}^*V_{a3}V_{b1}^*V_{b3}&F^C_n(a,k;b,l;i,j)
      =(V_{31}^*V_{33})^2[F^C_n(3,k;3,l;i,j)            &(9) \cr
 &+F^C_n(1,k;1,l;i,j) -F^C_n(1,k;3,l;i,j)-F^C_n(3,k;1,l;i,j)]. }
$$
Therefore, the effective lagrangian
$L^C_{\Delta B=2}$ is proportional to $(V_{31}^*V_{33})^2$.

     The large $t$-quark mass has two effects.
First, one of the $t$-squarks becomes light due to large
off-diagonal elements of the mass-squared matrix (2).
Second, the Yukawa interaction of $\tR$ with
the charged higgsino and the $b$-quark becomes
strong, so that the chargino coupling strengths with the
$t$-squarks are induced to be
different from those with the other up-type
squarks, as seen in eq. (8).  These effects soften the
cancellation among different squark contributions
in eq. (9), which otherwise is rather severe.

     The effective lagrangian (6) contains
two $\Delta B=2$ operators
$$
     O_V^{LL}=\bar d\rmuu\PL b\bar d\rmud\PL b, \quad
     O_S^{RR}=\bar d\PR b\bar d\PR b.
\eqno (10)
$$
The standard $W$-boson contribution
yields $O_V^{LL}$ alone, whereas
the chargino contribution yields both.  Therefore one
might expect that the SSM effects could be observed in
processes where $O_S^{RR}$ gives a sizable contribution.
For $\tb\sim 1$ the ratio $F_2^C/F_1^C$ is roughly estimated
as $(m_b/M_W)^2$, so that $O_S^{RR}$ can be neglected.
However, for a considerably large value of $\tb$
the effects of the operator $O_S^{RR}$ would become
sizable.

     For the charged Higgs boson contribution the exchanged
bosons in the box diagrams are either only charged Higgs
bosons or charged Higgs bosons and $W$-bosons.
The effective
lagrangian of the $\Delta B=2$ process is given by
$$\eqalign{
& L^H_{\Delta B=2} = {1\over 8M_W^2}({g^2\over 4\pi})^2
            \sum_{a,b}
                 V_{a1}^*V_{a3}V_{b1}^*V_{b3}  \cr
      & [F^H_1(a;b)\bar d\rmuu\PL b\bar d\rmud\PL b
         + F^H_2(a;b)\bar d\PR b\bar d\PR b],      }
\eqno(11)
$$
$$\eqalign{
F^H_1(a;b) &= {1\over 4\tan^4\beta}s_as_bY_1(r_H,r_H,s_a,s_b) \cr
 & + {1\over 2\tan^2\beta}s_as_bY_1(1,r_H,s_a,s_b)
   - {2\over \tan^2\beta}\sqrt{s_as_b}Y_2(1,r_H,s_a,s_b), \cr
F^H_2(a;b) &={m_b^2\over M_W^2}\sqrt{s_as_b}Y_2(r_H,r_H,s_a,s_b), \cr
 r_H &={\MCH^2\over M_W^2}, \quad
   s_a ={m_{ua}^2\over M_W^2}, }
$$
where $\MCH$ and $m_{ua}$ are respectively the charged
Higgs boson mass and the up-type quark mass of the $a$-th
generation.  Since the $t$-quark box diagrams predominate
over other diagrams, $L_{\Delta B=2}^H$
becomes approximately proportional to $(V_{31}^*V_{33})^2$.
The standard $W$-boson contribution is
also proportional to $(V_{31}^*V_{33})^2$, due to a predominant
$t$-quark box diagram.

     The effective lagrangians of the $\Delta S$=2 process
due to the chargino and the charged Higgs boson contributions
are obtained in a similar way.  The chargino-induced lagrangian
$L^C_{\Delta S=2}$ is shown to be proportional to
$(V_{31}^*V_{32})^2$.  The effects of the operator $O_S^{RR}$
are suppressed by $(m_s/M_W)^2$ and thus negligible.

     We now discuss how large the SSM contributions could be.
One observable for $\BB$ mixing is the mixing parameter
$x_d\equiv\DM_B/\Gamma_B$, $\DM_B$ and
$\Gamma_B$ being the mass difference and the average width
for the $B^0$-meson mass eigenstates.  The mass difference
is induced dominantly by the short distance contributions of
box diagrams.  Abbreviating the effective lagrangians as
$L_{\Delta B=2}=(1/8M_W^2)(g^2/4\pi)^2(V_{31}^*V_{33})^2AO_V^{LL}$,
we can express the mixing parameter $x_d$ as
$$
  x_d={G_F^2\over 6\pi^2}M_W^2{M_B\over \Gamma_B}
    f_B^2B_B|V_{31}^*V_{33}|^2|A_{tt}^W+A^C+A_{tt}^H|,
\eqno (12)
$$
where $G_F$, $f_B$, and $B_B$ are the Fermi constant,
the $B$-meson decay constant, and the bag factor.
The indices $W$, $C$, and $H$ stand for the
$W$-boson, the chargino, and the charged Higgs boson
contributions, respectively, and $tt$ for the
$t$-quark box diagrams.  The $O_S^{RR}$ operator
has been neglected.  The expression for $x_d$ in the
SM is given by eq. (12) with $A^C=A_{tt}^H=0$.

     The effects of the chargino and the charged Higgs boson
contributions are seen respectively by the ratios
$R_C=(A_{tt}^W+A^C)/A_{tt}^W$ and
$R_H=(A_{tt}^W+A_{tt}^H)/A_{tt}^W$.
In Fig. 1 we show $R_C$ as a function of the higgsino
mass parameter $\mH$ taking four sets of values for
$\mQ$ and $\tb$ listed in table 1.  The values of the
other parameters are set, as typical values,
for $\m2=200$ GeV, $\mQ=\mU=a_t\mgr$,
and $|c|=0.3$.  For the $t$-quark mass we use $m_t$=170 GeV.
Since QCD corrections in $A_{tt}^W$ and $A^C$ do not
differ much from each other, they almost cancel in
$R_C$ and can be neglected.  The mass of the lighter $t$-squark
corresponding to these parameter values are exhibited
in Fig. 2.  The mass of the lighter chargino is
smaller than 100 GeV for $-80$ GeV$\lsim \mH\lsim$160 GeV,
but the range $-20$ GeV$\lsim\mH\lsim$80 GeV
is experimentally ruled out since it leads to a chargino
lighter than 45 GeV.  We can see that the sign of
the chargino contribution is the same as that of the
$W$-boson contribution, and these contributions interfere
constructively.  In case (i.a) $R_C$ is larger
than 1.5 for $-90$ GeV$\lsim\mH\lsim$150 GeV and
$R_C\simeq 2.3$ in the region with the lighter
$t$-squark mass 45 GeV$\lsim{\tilde M}_{t1}\lsim$100 GeV.
The ratio $R_C$ can also have a value around 1.5 in
case (i.b).  The manifest dependence of $R_C$ on $\tb$
arises from the chargino Yukawa interactions:  $R_C$
increases as $\tb$ decreases, since a smaller value for
$v_2$ enhances the Yukawa couplings of the charginos to the
$t$-squarks.

     In Fig. 3 we show $R_H$
as a function of the charged Higgs Boson mass for (a)$\tb$=1.2
and (b)$\tb$=2.  The charged Higgs bosons also
contribute constructively.  The ratio $R_H$ is
larger than 1.5 for $\MCH\lsim$ 180 GeV in case (a).
Similarly to $R_C$, the value of $R_H$ increases
as $\tb$ decreases.  Note that all of
$A_{tt}^W$, $A^C$, and $A_{tt}^H$ have the same sign.
The physical effects of the SSM are given by their sum,
and the net contribution of the SSM is measured by
the ratio
$$
R={A_{tt}^W+A^C+A_{tt}^H\over A_{tt}^W}.
\eqno (13)
$$
This ratio becomes larger than $R_C$ and $R_H$ shown
in Fig. 1 and Fig. 3.  For example,
in case (i.a) with $\MCH=200$ GeV, the ratio $R$
becomes $R\simeq 1.9$ for $\mH=-100$ GeV
(${\tilde M}_{t1}\simeq 188$ GeV, $\mw1\simeq 119$ GeV)
and $R\simeq 2.7$ for $\mH=100$ GeV
(${\tilde M}_{t1}\simeq 85$ GeV, $\mw1\simeq 56$ GeV).

     For $\KK$ mixing the mass difference
$\DM_K$ receives large
long distance contributions, which have not yet been calculated
reliably.  As a result, it is not feasible to detect SSM
effects through $\DM_K$.  However, the $CP$ violation
parameter $\epsilon$ receives its dominant
contribution from the short distance effects, which
can be written as
$$\eqalign{
       \epsilon &= -{\rm e}^{i\pi/4}
     {G_F^2\over 12\r2\pi^2}M_W^2{M_K\over \Delta M_K}f_K^2B_K
      {\rm Im}[(V_{31}^*V_{32})^2(A_{tt}^W+A^C+A_{tt}^H)  \cr
       &+ (V_{21}^*V_{22})^2(A_{cc}^W+A_{cc}^H)
 + 2V_{31}^*V_{32}V_{21}^*V_{22}(A_{tc}^W+A_{tc}^H)].  }
\eqno (14)
$$
The term proportional to $(V_{31}^*V_{32})^2$ is enhanced
by the same amount as for $\BB$ mixing given by eq. (13),
whereas the enhancements of the other terms due to the charged
Higgs boson contribution are small.

     We next discuss the implication of the above
enhancements in $x_d$ and $\epsilon$ for the
evaluation of the CKM matrix, and consider possible
constraints on the SSM parameters.
In Fig. 4 we show the allowed range for
the ratio $R$ derived from
the experimental values of $x_d$, $\epsilon$, and
CKM matrix elements, as a function
of $\cos\delta$, $\delta$ being the $CP$-violating phase
appearing in the standard parametrization [7] of
the CKM matrix.  As a typical example, we have taken
$|V_{12}|=0.22$, $|V_{23}|=0.4$, $|V_{13}/V_{23}|=0.08$ [10],
$x_d=0.71 [11], |\epsilon|=2.26\times 10^{-3}$ [7] and
incorporated the uncertainties
of $B_K$, $B_B$, and $f_B$ as 0.6 $<B_K<$ 0.9 [12] and
180 MeV $<\sqrt{f_B^2B_B}<$ 260 MeV [13].
For the QCD correction
factors we have used 0.55 for $A_{tt}^W$ in $\BB$ mixing and
0.57, 1.1, and 0.36 for $A_{tt}^W$, $A_{cc}^W$, and $A_{tc}^W$
in $\KK$ mixing [14].
The regions between the
solid curves and between the dashed curves are respectively
allowed by $x_d$ and $\epsilon$.
In the region consistent with both $x_d$
and $\epsilon$, the ratio $R$ is smaller than or around 2.
The SSM with $R>1$ favors
for $\cos\delta$ a value larger than the one predicted by the
SM ($R=1$).
Although present uncertainties
in the input parameters make it difficult to state a definite
prediction, the SSM parameter regions which give
$R>2$ are likely to be in conflict with the experimental values
of the CKM matrix.

     In conclusion, we have studied $\BB$ and $\KK$ mixings within
the framework of the SSM, emphasizing
the importance of the box diagrams mediated
by charginos and charged Higgs bosons.
If a chargino, a $t$-squark, and/or a charged Higgs boson
have masses around 100 GeV or less and $\tb$ is not much
larger than unity, the SSM contribution to
the mixing parameter $x_d$ could be enhanced by a factor
around 2, compared to the SM contribution.  The SSM
contribution to the $CP$ violation parameter $\epsilon$
could also be enhanced by a comparable amount.  Note that
although the enhancements of $\epsilon$ and $x_d$ are
correlated, they are not equal due to the presence
of the last two terms in eq. (14) whose contribution
to $\epsilon$ cannot be neglected.

     Experimentally, fairly precise measurements have been
achieved for $x_d$ [11] and $\epsilon$ [7].
The experimental values of these quantities can be
used to determine the CKM matrix elements.
If the ratio $R$ defined in eq. (13) is
larger than unity, the values of $|V_{31}^*V_{33}|^2$ and
Im[$(V_{31}^*V_{32})^2$] are predicted to be smaller
than those obtained in the SM.  On the other hand,
some of the CKM matrix elements have been
measured by particle decays for which
new contributions by the SSM
are negligible.  Owing to the unitarity of the
CKM matrix, all of  these measured values are not
independent of each other.
We have shown that the present knowledge for the CKM
matrix could already give nontrivial constraints on
the SSM.
Future measurements will give
further information on the CKM matrix.  For
example, the measurement of $CP$ asymmetries (for
reviews, see ref. [15]) in $B^0$ decays will
determine the angles of the unitarity triangle
formed by $V_{31}V_{33}^*$, $V_{21}V_{23}^*$, and
$V_{11}V_{13}^*$.  The study of the compatibility
of all these measurements, taking into account
the stringent constraints of CKM unitarity, has the
potential of either providing indirect evidence for
the SSM or putting stringent bonds on some of
its parameters.

\medskip
     N.O. would like to thank S. Komamiya for discussions on
experimental status of $\BB$ mixing at LEP.   G.C.B. thanks
the CERN Theory Division for their hospitality.  The work
of G.C.B. was supported in part by Science Project No.
SCI-CT91-0729 and EC contract No. CHRX-CT93-0132.  This work
is also supported in part by Grant-in-Aid for Scientific Research
(G.~C.~Cho) and (No. 06640409, Y.~K.) from the Japanese Ministry
of Education, Science and Culture.

\vfill \eject

{\bf References}
\smallskip
\item{[1]} J. Ellis and D.V. Nanopoulos, Phys. Lett. 110B
           (1982) 44;
\item{}    R. Barbieri and R. Gatto, Phys. Lett. 110B
           (1982) 211;
\item{}    T. Inami and C.S. Lim, Nucl. Phys. B207 (1982)
           533;
\item{}    M.J. Duncan, Nucl. Phys. B221 (1983) 285;
\item{}    J.F. Donoghue, H.P. Nilles, and D. Wyler,
           Phys. Lett. 128B (1983) 55;
\item{}    A. Bouquet, J. Kaplan, and C.A. Savoy, Phys. Lett.
           148B (1984) 69.
\item{[2]} L.F. Abbott, P. Sikivie, and M.B. Wise,
           Phys. Rev. D21 (1980) 1393.
\item{[3]} W. Grimus, Fortschr. Phys. 36 (1988) 201.
\item{[4]} T. Kurimoto, Phys. Rev. D39 (1989) 3447;
\item{}    S. Bertolini, F. Borzumati, A. Masiero, and G. Ridolfi,
           Nucl. Phys. B353 (1991) 591;
\item{}    F.M. Borzumati, DESY-93-090.
\item{[5]} CDF Collaboration, FERMILAB-PUB-94/097-E;
           FERMILAB-PUB-94/116-E.
\item{[6]} H. Baer, M. Drees, R. Godbole, J.F. Gunion, and X. Tata,
           Phys. Rev. D44 (1991) 725.
\item{[7]} Review of Particle Properties, Phys. Rev. D45 (1992) S1.
\item{[8]} N. Oshimo, Nucl. Phys. B404 (1993) 20.
\item{[9]} H.P. Nilles, Phys. Rep. 110 (1984) 1;
\item{}    P. Nath, R. Arnowitt, and A.H. Chamseddine, {\it Applied
           N=1 Supergravity}
\item{}    (World Scientific, Singapore, 1984);
\item{}    H.E. Haber and G.L. Kane, Phys. Rep. 117 (1985) 75.
\item{[10]}D.Z. Besson, in Proc. of the XVI International Symposium
           on Lepton and Photon Interactions, Ithaca, 1993.
\item{[11]}CLEO Collaboration, Phys. Rev. Lett. 71 (1993) 1680;
\item{}    ARGUS Collaboration, Phys. Lett. B324 (1994) 249;
\item{}    S. Tarem, talk given at the Rencontre de Moriond, Les
           Arcs, 1994.
\item{[12]}P.B. Mackenzie, in Proc. of the XVI International
           Symposium on Lepton and Photon Interactions, Ithaca,
           1993.
\item{[13]}A. Abada, LPTHE Orsay-94/57.
\item{[14]}A.J. Buras, M.E. Lautenbacher, and G. Ostermaier,
           MPI-Ph/94-14.
\item{[15]}I.I. Bigi, V.A. Khoze, N.G. Uraltsev, and A.I.
           Sanda, in {\it CP violation}, ed. C. Jarlskog
           (World Scientific, Singapore, 1989);
\item{}    Y. Nir and H.R. Quinn, SLAC-PUB-5737, 1992.

\vfill\eject

{\bf Table Caption}
\bigskip
\item{Table 1:} The values for $\mQ$ and $\tb$.

\vskip 2.0 true cm
{\bf Figure Captions}
\bigskip
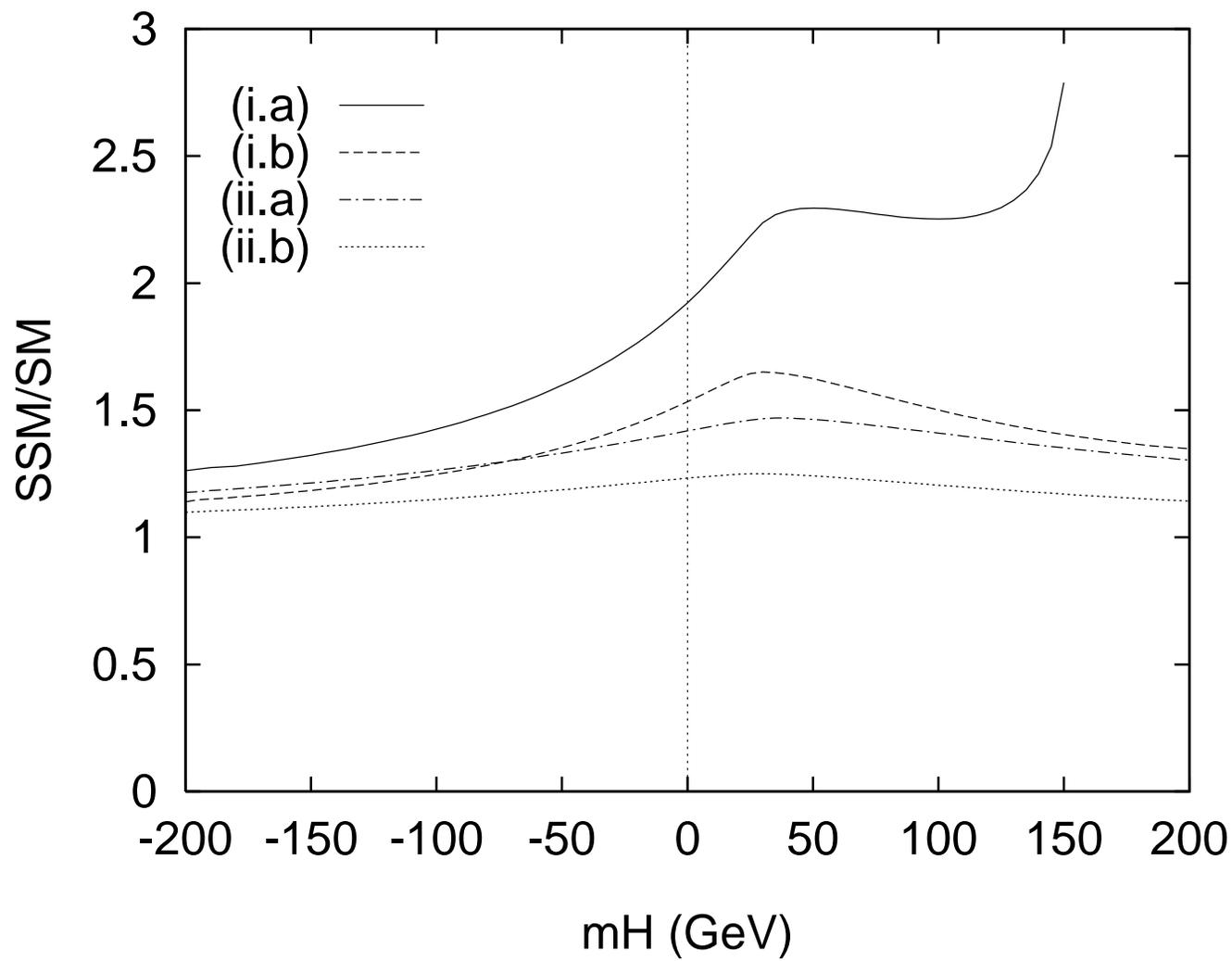
\item{Fig. 1:} The ratio $R_C$ for the parameter values in table 1.
               Other parameters are fixed as $\m2$=200 GeV,
               $\mQ=\mU=a_t\mgr$, and $|c|$=0.3.
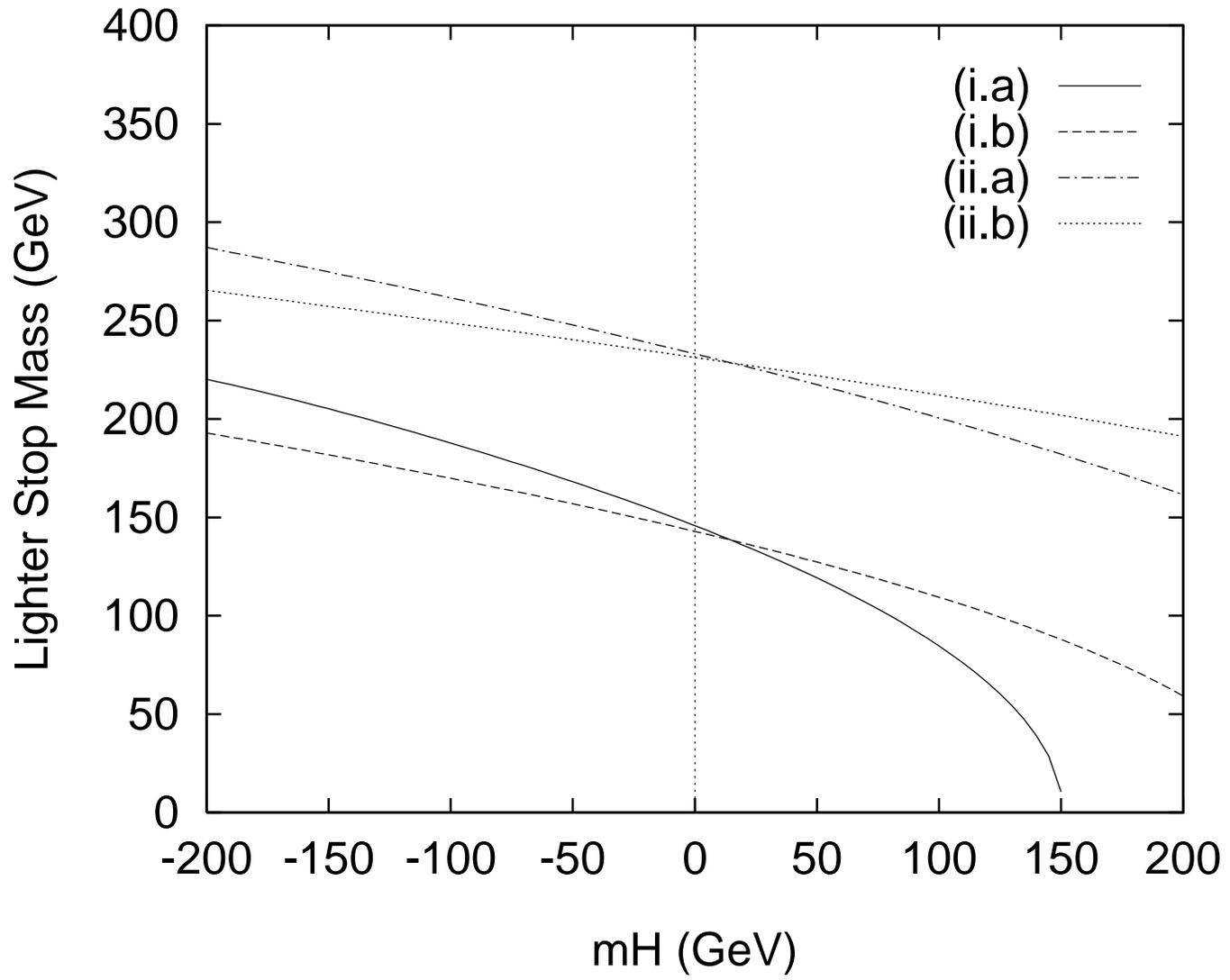
\item{Fig. 2:} The lighter $t$-squark mass for the same parameter
               values as in Fig. 1.
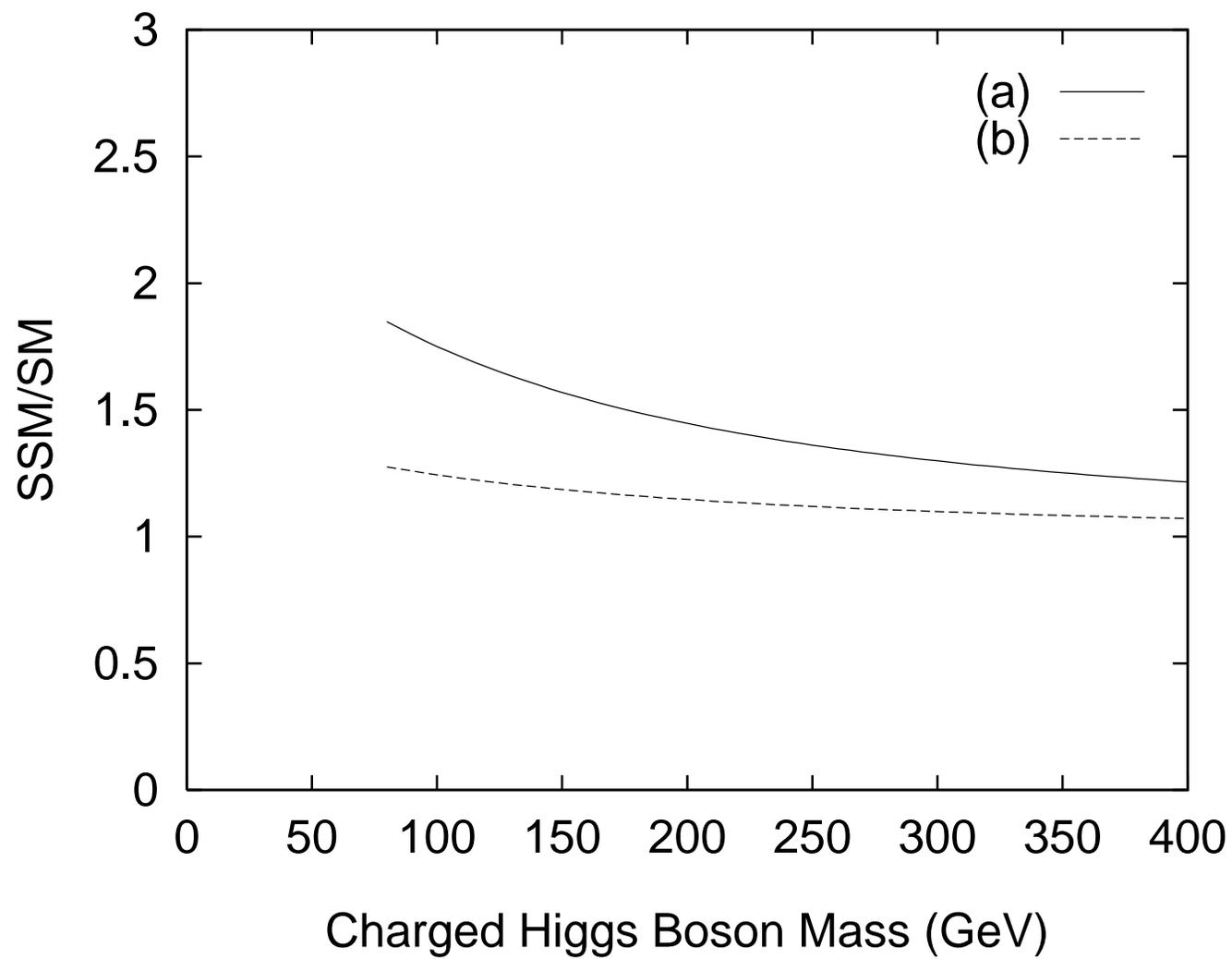
\item{Fig. 3:} The ratio $R_H$ for (a)$\tb$=1.2 and (b)$\tb$=2.
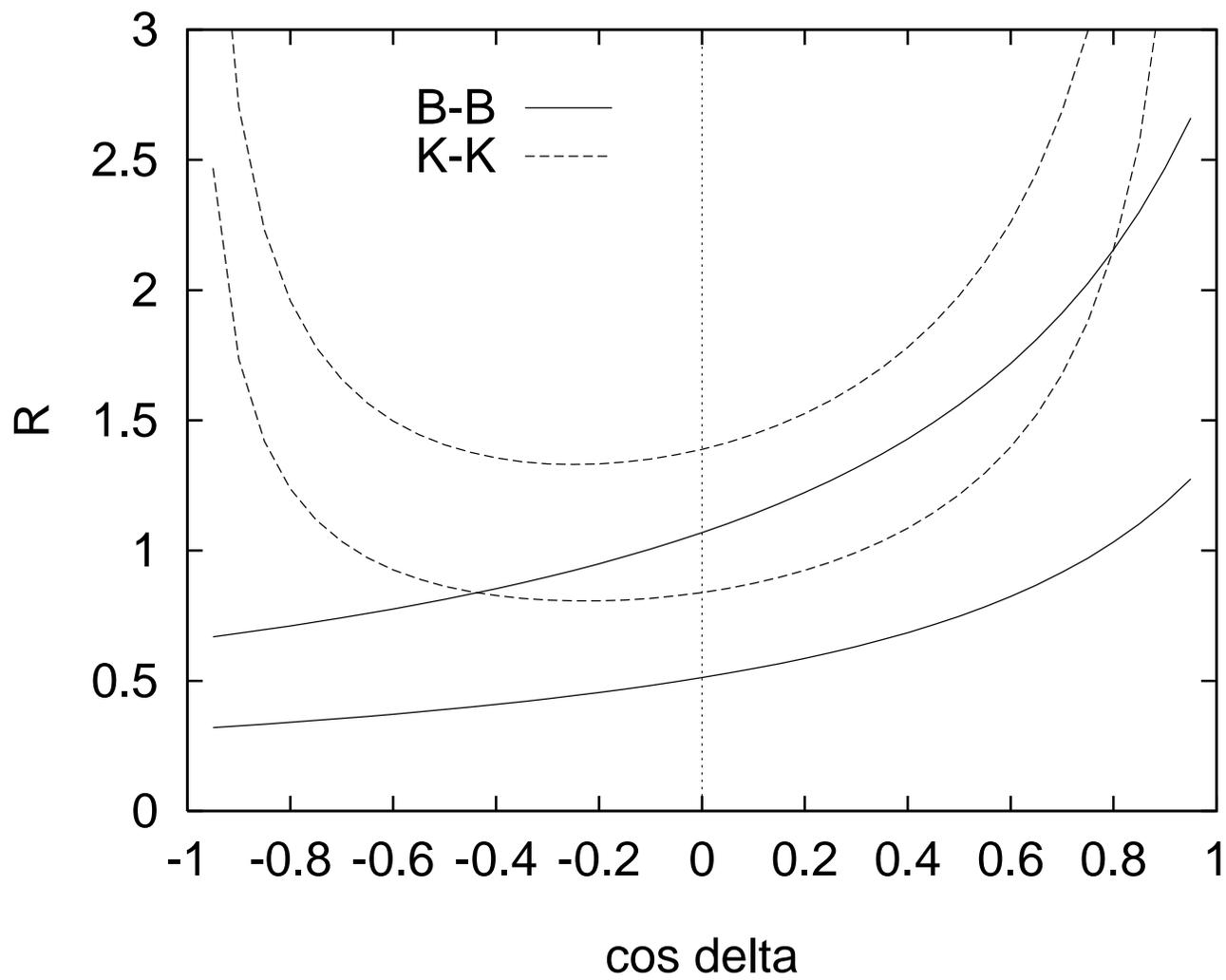
\item{Fig. 4:} The ratio $R$ allowed by the experimental values
               of $x_d$ and $\epsilon$.

\vskip 4.0 true cm
$$\vbox{\offinterlineskip
\hrule
\halign{&\vrule#&
         \strut\quad\hfil#\hfil\quad\cr
height2pt&\omit&&\omit&&\omit&&\omit&&\omit&\cr
&   && (i.a)&& (i.b) &&(ii.a)&&(ii.b)&\cr
height2pt&\omit&&\omit&&\omit&&\omit&&\omit&\cr
\noalign{\hrule}
height2pt&\omit&&\omit&&\omit&&\omit&&\omit&\cr
&$\mQ$ (GeV) &&200&&200&&300&&300&\cr
height2pt&\omit&&\omit&&\omit&&\omit&&\omit&\cr
\noalign{\hrule}
height2pt&\omit&&\omit&&\omit&&\omit&&\omit&\cr
&$\tb$ &&1.2&&2.0&&1.2&&2.0&\cr
height2pt&\omit&&\omit&&\omit&&\omit&&\omit&\cr}
\hrule
}$$
\vskip 0.5 true cm
\centerline{Table 1}
\vfill \eject

\end